\documentclass[aps,prd,onecolumn,12pt,superscriptaddress,nofootinbib,nobibnotes,longbibliography]{revtex4-1}

\usepackage{amssymb}
\usepackage{graphicx}
\usepackage{amsmath}
\usepackage{hyperref}
\usepackage{subfigure}
\usepackage{multirow}
\usepackage{setspace}
\usepackage{verbatim}
\usepackage{float}
\usepackage{color}
\usepackage{ulem}
\usepackage[utf8]{inputenc}
\usepackage[table,xcdraw]{xcolor}
\usepackage{makecell}
\usepackage{url}
\usepackage{bm}
\usepackage{placeins}

\newcommand{\red}{\textcolor{black}}

\begin{document}


\title{Prospect on constraining environment-dependent dilaton model from gravitational redshift measurements}

\author{Li Hu}
\email{huli21@mails.ucas.ac.cn}
\affiliation{School of Fundamental Physics and Mathematical Sciences, Hangzhou Institute for Advanced Study (HIAS), University of Chinese Academy of Sciences (UCAS), Hangzhou 310024, China}
\affiliation{Institute of Theoretical Physics, Chinese Academy of Sciences (CAS), Beijing 100190, China}
\affiliation{University of Chinese Academy of Sciences (UCAS), Beijing 100049, China}

\author{Rong-Gen Cai}
\email{caironggen@nbu.edu.cn}
\affiliation{Institute of Fundamental Physics and Quantum Technology, \& School of Physical Science and Technology, Ningbo University, Ningbo, 315211, China}

\author{Song He}
\email{hesong@nbu.edu.cn}
\affiliation{Institute of Fundamental Physics and Quantum Technology, \& School of Physical Science and Technology, Ningbo University, Ningbo, 315211, China}
\affiliation{Max Planck Institute for Gravitational Physics (Albert Einstein Institute), Am M\"{u}hlenberg 1, 14476 Golm, Germany}

\author{Li-Fang Li}
\email{lilifang@imech.ac.cn}
\affiliation{Center for Gravitational Wave Experiment, National Microgravity Laboratory, Institute of Mechanics, Chinese Academy of Sciences, Beijing 100190, China}

\author{Tong Liu}
\email{liutong2021@csu.ac.cn}
\affiliation{Key Laboratory Of Space Utilization, Technology And Engineering Center For Space Utilization, Chinese Academy Of Sciences, Beijing 100094, China}

\author{Peng Xu}
\email{xupeng@imech.ac.cn}
\affiliation{Center for Gravitational Wave Experiment, National Microgravity Laboratory, Institute of Mechanics, Chinese Academy of Sciences, Beijing 100190, China}
\affiliation{Taiji Laboratory for Gravitational Wave Universe (Beijing/Hangzhou), University of Chinese Academy of Sciences (UCAS), Beijing 100049, China}
\affiliation{Key Laboratory of Gravitational Wave Precision Measurement of Zhejiang Province, Hangzhou Institute for Advanced Study, UCAS, Hangzhou, 310024, China}
\affiliation{Lanzhou Center of Theoretical Physics, Lanzhou University, Lanzhou 730000, China}

\author{Shao-Jiang Wang}
\email{schwang@itp.ac.cn (Corresponding author)}
\affiliation{Institute of Theoretical Physics, Chinese Academy of Sciences (CAS), Beijing 100190, China}
\affiliation{Asia Pacific Center for Theoretical Physics (APCTP), Pohang 37673, Korea}

\begin{abstract}
Scalar-tensor gravity represents a natural extension of general relativity. This paper investigates \red{the prospects for constraining the environmentally dependent dilaton through atomic-clock comparisons across media of different densities. Using the scalar Compton wavelength, we assess how the finite size of the surrounding medium and screening by the clock enclosure limit the accessible region. To test the continuous-density approximation for dilute media, we model matter as a periodic lattice of spherical particles and compare an analytic Neumann solution with a three-dimensional finite-element solution on the cubic box. The continuous and discrete models give indistinguishable redshift predictions near the $10^{-20}$ and $10^{-25}$ benchmarks. In the strong-coupling regime, differences become apparent on redshift contours of order $10^{-46}$ or lower. Under the assumed environmental and instrumental conditions, comparisons involving a dense medium identify a window for probing smaller values of $A_2$, offering a gravitational-redshift test complementary to existing fifth-force searches.}
\end{abstract}
\maketitle

\section{Introduction} 

Recently, Levy and Uzan have proposed a new probing method, gravitational redshift~\cite{Levy:2024vyd}, to constrain screened scalar-tensor gravity theory. The gravitational redshift experiment is a classical test method based on potential effects. Due to the remarkable precision of atomic clocks, gravitational redshift experiments are expected to test gravitational theories with much higher precision than many other existing methods.

The scalar-tensor gravity theory extends the framework of general relativity by introducing a dynamical scalar field. It provides new perspectives on the early Universe~\cite{Capozziello:2011et} and dark energy~\cite{Elizalde:2004mq, Abdalla:2022yfr}. Meanwhile, to pass the existing tests at sub-solar scales, mainly the fifth force tests~\cite{Kapner:2006si, Brax:2022uiv, Brax:2022uyh}, gravity theories with screening mechanisms should be given priority. The so-called screening mechanisms allow for order-one (or larger) deviations from general relativity on large scales while suppressing these modifications on smaller scales~\cite{Sakstein:2014jrq}. This type of method includes chameleon ~\cite{Khoury:2003rn}, symmetron ~\cite{Hinterbichler:2010es}, Damour–Polyakov effect ~\cite{Damour:1994zq}, Vainshtein mechanism ~\cite{Vainshtein:1972sx}, along with others~\cite{Khoury:2025txd}.

Testing models with screening mechanisms through experiments has long been a fascinating topic~\cite{Upadhye:2012rc, Xu:2014wda, Sabulsky:2018jma, Brax:2010gp, Jenke:2014yel, Fischer:2023eww, Fischer:2023koa}. Traditional tests can be broadly categorized into those based on the fifth force and those based on potential effects. Furthermore, depending on whether the probe involves a quantum-mechanical description, these tests can also be classified as classical or quantum probes.

In scalar-tensor gravity theory, it is essential to have an appropriate modeling for matter distributions in space since the reading of an atomic clock is affected by scalar field in that space, and scalar field itself also depends on matter distributions. In this case, there is a crucial question: if the continuous fluid approximation is still valid for low-density environments? \red{This problem is very similar to the average problem in cosmology and can be overcame by treating it as a lattice.~\cite{10.1111/j.1365-2966.2012.21750.x, Fleury:2016tsz, Fleury:2018cro, Bruneton:2012ru, Uzan:2025mjz}.} In low-density environments, all mass is concentrated in microscopic particles such as atoms or molecules, and these particles are so far apart from each other. If the average density of space is obtained first, and then one uses the average density to calculate the associated scalar field distribution, the results may differ significantly from the modeling with a discrete mass distribution \red{in certain parameter regions}. 

In this paper, we concentrate on a specific example of the Damour-Polyakov effect, environment-dependent dilaton~\cite{Brax:2010gi}. See~\cite{Fischer:2024eic} for a recent review. Its approach to circumventing the fifth force constraints differs from Ref.~\cite{Gasperini:2001pc}, which requires the Damour-Polyakov mechanism with a minimum at infinity since it assumes that the string scale is at the same order of magnitude as the Planck scale. Instead, our environment-dependent approach assumes that the string scale is much lower than the Planck scale, allowing the coupling to vanish for a finite value of the dilaton. This mechanism is not only a part of the low-energy effective description of string theory, but also plays a crucial role in understanding the structure, symmetry, and non-perturbative properties of string theory itself. In terms of gravitational redshift, the advantage of this model lies in the fact that, after appropriate simplification, the corresponding field equation is linear, which enables us to solve the equation of motion by analytical methods to gain a precise understanding of the scalar field.


It is important to note that most existing experimental constraints on environment-dependent dilaton are primarily sensitive to relatively strong couplings~\cite{Fischer:2023koa, Fischer:2023eww, Feleppa:2025vop}. Consequently, a significant portion of the parameter space, particularly in the weak-coupling regime, remains unconstrained. This motivates the exploration of alternative probes, such as gravitational redshift experiments, which can access complementary regions of parameter space.


\red{To achieve the goal of constraining the parameter space of the dilaton with gravitational redshift, we first calculate the scalar contribution for idealized density pairs and assess how the size of the surrounding medium and the thickness of the clock enclosure limit the applicability of these predictions. We then use a periodic lattice of spherical particles to test the continuous-density approximation for dilute media. To assess the spherical-cell approximation, we compare the analytic Neumann solution with a three-dimensional finite-element calculation on the cubic box. The continuous and discrete models yield indistinguishable redshift predictions near the clock benchmarks within the resolution of our comparison, with appreciable departures emerging at stronger coupling. These results identify the conditions under which clock comparisons involving a dense medium could probe smaller values of $A_2$.}

\section{Conformal scalar-tensor gravity}

In this paper, we work in natural units with $\hbar=c=1$. Meanwhile, the reduced Planck mass is $M_\mathrm{Pl} = \sqrt{1/8\pi G}$. Unless specified, the range of the Greek index takes $0\sim3$, and the range of the Latin index takes $1\sim3$. The Minkowski spacetime metric is $\eta_{\mu\nu}=\mathrm{diag}(-,+,+,+)$. In curved spacetime, we keep the signature convention of the metric consistent with Minkowski spacetime. 

\subsection{Environment-dependent dilaton}

In the strong coupling regime, the coupling between the dilaton and gravity has a low-energy effective action~\cite{Damour:1994zq}. In the Jordan frame, this effective action can be expressed as~\cite{Brax:2010gi}
\begin{align}
S&=\int\sqrt{-\tilde{g}}\,\mathrm{d}^4 x\Big[\frac{e^{-2\psi(\phi)}}{2l_s^2}\tilde R+\frac{Z(\phi)}{2l_s^2}(\tilde\nabla\phi)^2-\tilde V(\phi)\Big]+S_\mathrm{mat}(\tilde g_{\mu\nu},\Psi_i;g_i(\phi))\,,
\end{align}
where $\Psi_i$ represents a matter field, the conformal metric $\tilde g_{\mu\nu}=A^2(\phi)g_{\mu\nu}$ involves a  dilaton coupling $A(\phi)=M_\mathrm{Pl}l_s e^{\psi(\phi)}$ as the conformal factor, the string length scale $l_s$ and dilaton coupling $\psi(\phi)$ will not be given explicitly. Its rescaling freedom is fixed by requiring $A(\phi_*)=1$, where $\phi_*$ is approximately today's value. Note that the coupling ``constant" $g_i(\phi)$ here is scalar-field dependent, different from either chameleon or symmetron mechanisms.

If one converts the action to the Einstein frame, it is not difficult to find that
\begin{align}
S&=\int\sqrt{-{g}}\,\mathrm{d}^4 x\Big[\frac{M^2_\mathrm{pl}}{2} R-\frac{k^2(\phi)}{2}(\nabla\phi)^2- V(\phi)\Big]+S_\mathrm{mat}(A^2(\phi)g_{\mu\nu},\Psi_i;g_i(\phi))\,,
\end{align}
where $k^2(\phi)=6\beta^2(\phi)-A^2(\phi)Z(\phi)/l_s^2$ with $ \beta(\phi)=M_\mathrm{Pl}\mathrm{d}\ln A/\mathrm{d}\phi$, and $V(\phi)=A^4\tilde V(\phi)$. In Ref.~\cite{Brax:2010gi}, the authors further assume that today we are in the strong coupling limit, which means $e^{-\phi_*/M_\mathrm{Pl}}\ll1$. In this limit ($\phi\rightarrow\infty$), one could further assume that~\cite{Gasperini:2001pc, Brax:2010gi}
\begin{align}
\tilde V(\phi)&=V_1 e^{-\frac{\phi}{M_\mathrm{Pl}}}+\mathcal{O}(e^{-\frac{2\phi}{M_\mathrm{Pl}}})\,,\\
Z(\phi)&=-\frac{l_s^2}{\lambda^2}+b_z e^{-\frac{\phi}{M_\mathrm{Pl}}}+\mathcal{O}(e^{-\frac{2\phi}{M_\mathrm{Pl}}})\,,\\
g_i^{-2}(\phi)&=\bar{g}^{-2}+b_i e^{-\frac{\phi}{M_\mathrm{Pl}}}+\mathcal{O}(e^{-\frac{2\phi}{M_\mathrm{Pl}}})\,,
\end{align}
where $b_z$ and $b_i$ are $\mathcal{O}(1)$ constants. In this case, $k(\phi)$ can be further simplified to $k(\phi)=\frac{1}{\lambda}\sqrt{1+6\lambda^2\beta^2(\phi)}$.

Now that one has gained the simplified version of the complete action in the Einstein frame, the next step is to obtain equations of motion (EOM) as follows:
\begin{align}
G_{\mu\nu}&\equiv R_{\mu\nu}-\frac{1}{2}Rg_{\mu\nu}=\frac{1}{M^2_\mathrm{pl}}(T_{\mu\nu}+T^{(\phi)}_{\mu\nu})\,,\label{EOM1}\\
\square\hat{\phi}&=\frac{\mathrm{d}V}{\mathrm{d}\hat{\phi}}-\frac{\mathrm{d}\ln A}{\mathrm{d}\hat{\phi}}T\red{-\frac{S_i}{2k(\phi)M_\mathrm{Pl}\sqrt{-g}}g_i^2(\phi)b_ie^{-\frac{\phi}{M_\mathrm{Pl}}}\,.}\label{EOM2}
\end{align}
In the above EOM, the stress-energy tensor takes the form $T_{\mu\nu}\equiv(-2/\sqrt{-g})(\delta S_\mathrm{mat}/\delta g^{\mu\nu})$, with its trace given by $T=g^{\mu\nu}T_{\mu\nu}$. 
Moreover, $S_i=\delta S_\mathrm{mat}/\delta\ln g_i$. \red{Here, $g_i(\phi)$ denote the dilaton-dependent gauge coupling functions and $\delta\ln g_i/\delta\phi\sim\frac12M_\mathrm{Pl}^{-1}g_i^2(\phi)b_ie^{-\phi/M_\mathrm{Pl}}$ is used.}
Since $S_i$ is of $\mathcal{O}(T)$, and $e^{-\phi/M_\mathrm{Pl}}\ll1$, one can safely discard the last term. Meanwhile, $\mathrm{d}\hat{\phi}=k(\phi)\mathrm{d}\phi$, assuming $k(\phi)\equiv1$ is a reasonable simplification~\cite{Brax:2017wcj}. Using the fact that $T=-\rho$ for non-relativistic matter, the EOM of scalar field Eq.~(\ref{EOM2}) is simplified into
\begin{align}
\square{\phi}&\equiv g^{\mu\nu}\nabla_{\mu}\nabla_{\nu}{\phi}=\frac{\mathrm{d}V}{\mathrm{d}{\phi}}+\frac{\mathrm{d}\ln A}{\mathrm{d}{\phi}}\rho\,.
\end{align}

Then the above equation can be integrated to find the effective potential for $\phi$, which is
\begin{align}
    V_\mathrm{eff}(\phi)&=V(\phi)+\rho[A(\phi)-1]=V_1A(\phi)^4e^{-\frac{\phi}{M_\mathrm{Pl}}}+\rho[A(\phi)-1]\,.
\end{align}
Here, $A(\phi)$ close to unity is used again. So far, the functional form of $A(\phi)$ is unknown, but the Damour-Polyakov mechanism requires that $A(\phi)$ has a minimum at some $\phi=\phi_\ast$, thus, Ref.~\cite{Brax:2010gi} assumes that the function has the following expansion form,
\begin{align}
    A(\phi)=1+\omega(\phi)=1+\frac{A_2}{2M^2_\mathrm{pl}}(\phi-\phi_\ast)^2\,,
\end{align}
\red{with $|w(\phi)|\ll1$. Then, we work in the regime where the dimensionless parameter $A_2\gg1$. Introducing a dimensionless variable $\gamma=(\phi-\phi_\ast)/M_\mathrm{Pl}$ and defining $V_0\equiv V_1e^{-\phi_\ast/M_\mathrm{Pl}}$, we expand the effective potential through second order in $\gamma$,}
\begin{align}
    \red{V_\mathrm{eff}(\phi)}&\red{=V_0A(\phi)^4e^{-\gamma}+\rho\big[A(\phi)-1\big]}\nonumber\\
    &\red{= V_0\big[1+2A_2\gamma^2+\mathcal{O}(\gamma^4)\big]\big[1-\gamma+\frac{1}{2}\gamma^2+\mathcal{O}(\gamma^3)\big]+\frac{1}{2}\rho A_2\gamma^2}\nonumber\\
    &\red{= V_0-V_0\gamma+\frac{1}{2}\big[A_2(\rho+4V_0)+V_0\big]\gamma^2+\mathcal{O}(\gamma^3)}\nonumber\\
    &\red{\approx V_0-V_0\gamma+\frac{1}{2}\big[A_2(\rho+4V_0)\big]\gamma^2}\,,
\end{align}
\red{where the requirement $A_2\gg1$ is used in the last step. Within this approximation, the minimum is located at}
\begin{align}
    \phi_\mathrm{min}-\phi_{\ast}=\frac{V_0M_\mathrm{Pl}}{A_2(4V_0+\rho)}\,.
\end{align}
\red{In terms of the displacement $\varphi\equiv\phi-\phi_{\ast}$, the approximate potential reads}
\begin{align}
    V_\mathrm{eff}(\varphi)= V_0-V_0\bigg(\frac{\varphi}{M_\mathrm{Pl}}\bigg)+\frac{A_2}{2}(\rho+4V_0)\bigg(\frac{\varphi}{M_\mathrm{Pl}}\bigg)^2\,.
\end{align}
The field value that minimizes the effective potential is
\begin{align}\label{eq:extremum}
    \varphi_\mathrm{min}=\frac{V_0M_\mathrm{Pl}}{A_2(4V_0+\rho)}\,.
\end{align}
In addition, the effective mass takes $m^2_{\varphi}(\rho)\equiv\mathrm{d}^2V_\mathrm{eff}/\mathrm{d}\varphi^2=A_2(\rho+4V_0)/M_\mathrm{Pl}^2$ and the corresponding effective Compton wavelength of the scalar field is 
\begin{align}
    \lambda_{\varphi}\equiv1/m_{\varphi}=\frac{M_\mathrm{Pl}}{\sqrt{A_2(4V_0+\rho)}}\,.
\end{align}

Before discussing the Newtonian limit of the environmentally dependent dilaton model, it is useful to briefly review the existing experimental constraints on this class of models. In recent years, several studies have placed bounds on the parameter space of the dilaton model. For instance, analyses based on qBounce experiments, Lunar Laser Ranging, and Casimir And Non Newtonian force EXperiment indicate that values of $A_2$ in the range $10^{30}\sim 10^{50}$ are excluded~\cite{Fischer:2023koa}. Furthermore, neutron interferometry and measurements based on the LAGEOS-2 satellite have placed constraints excluding a similar range of $A_2$~\cite{Fischer:2023eww, Feleppa:2025vop}.

It should be emphasized that most of these constraints arise from tests based on the fifth force. In such experiments, generating observable signals typically requires a sufficiently strong coupling between the scalar field and matter, corresponding to relatively large values of $A_2$. In contrast, gravitational redshift experiments based on potential effects are more sensitive in the region $A_2<10^{30}$ \red{(See Fig.~\ref{fig:average})}. Therefore, gravitational redshift experiments provide a genuinely complementary probe of the parameter space, accessing regions that are largely unconstrained by existing fifth-force experiments.

\subsection{Newtonian limit}

Since the existing technology generally places atomic clocks in the Earth's gravitational field, it is reasonable to choose a static Newtonian metric as the background in the Einstein frame,
\begin{align}
    \mathrm{d}s^2=-(1+2\Phi)\mathrm{d}t^2+(1-2\Phi)(\mathrm{d}x^2+\mathrm{d}y^2+\mathrm{d}z^2)\,.
\end{align}
The 00-component of Eq.~(\ref{EOM1}) and (\ref{EOM2}) can be described as    
\begin{align}
    2M^2_\mathrm{pl}\Delta\Phi&=\rho-2V(\varphi)\,,\\
    \Delta\varphi=\frac{\mathrm{d}V_\mathrm{eff}(\varphi)}{\mathrm{d}\varphi}&=-\frac{V_0}{M_\mathrm{Pl}}+\frac{A_2(\rho+4V_0)}{M_\mathrm{Pl}^2}\varphi\,.
\end{align}

In the Jordan frame, the conformal metric $\tilde g_{\mu\nu}$ can also be expanded about the Minkowski metric $\eta_{\mu\nu}$, which is $\tilde g_{\mu\nu}=\eta_{\mu\nu}+\tilde h_{\mu\nu}$. The line element is expected to be expressed as follows due to the existence of gauge freedom,
\begin{align}
    \mathrm{d}\tilde s^2=-(1+2\tilde\Phi)\mathrm{d}t^2+(1-2\tilde\Phi)(\mathrm{d}x^2+\mathrm{d}y^2+\mathrm{d}z^2)\,.
\end{align}

\begin{figure}[h]
    \centering
    \includegraphics[width=0.5\textwidth]{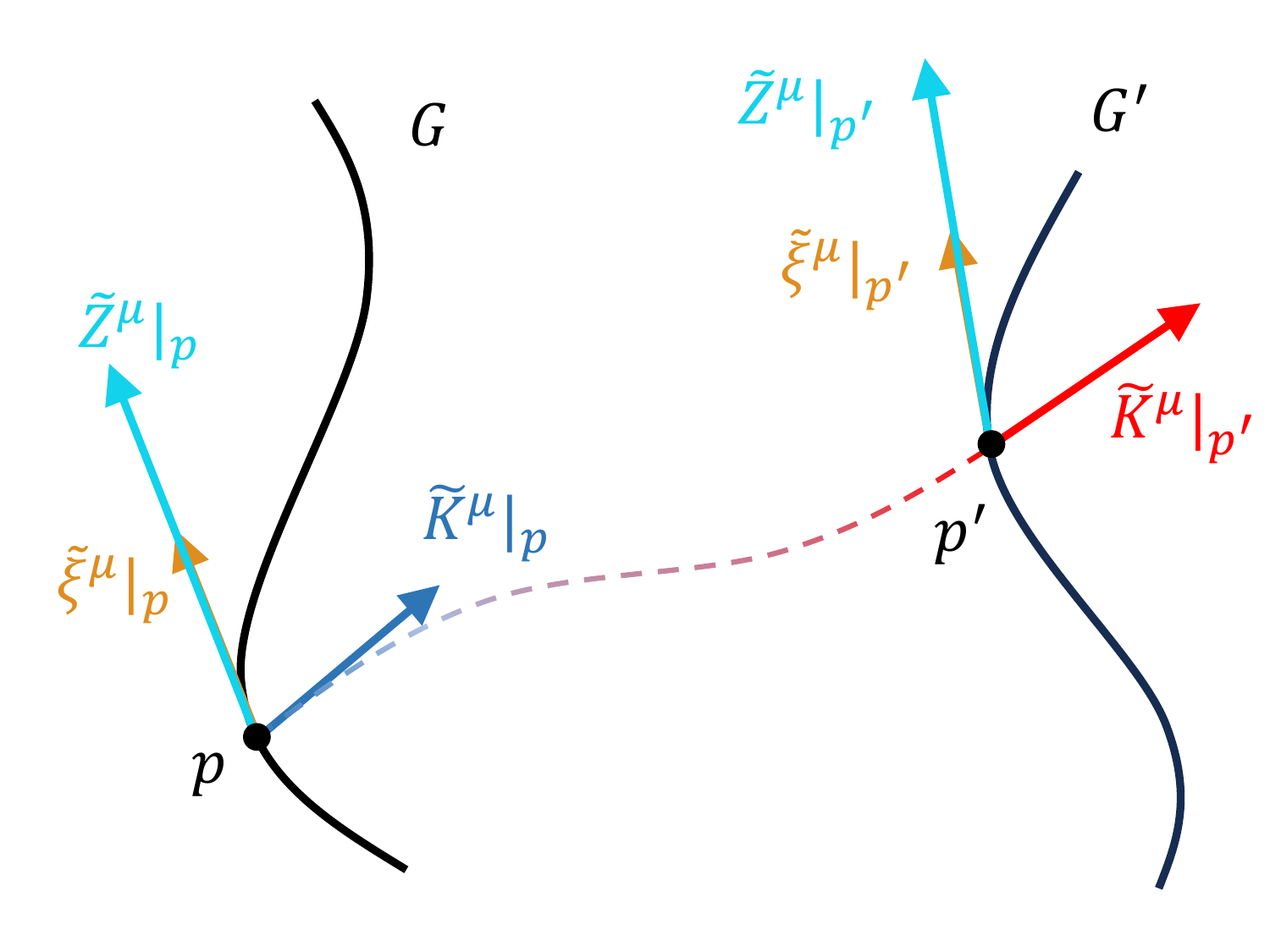}
	\caption{Schematic diagram of gravitational redshift.}
    \label{fig:sketch}
\end{figure}

Since the metric and the matter field are minimally coupled in the Jordan frame, particles follow geodesics. 
If the light emitted from point $p$ in spacetime is received at point $p'$, the corresponding gravitational redshift can be expressed as~\cite{Liang:2023ahd}
\begin{align}
    1+z\equiv\frac{E|_p}{E|_{p'}}\,.
\end{align}
Fig.~\ref{fig:sketch} is a schematic diagram of gravitational redshift in a stationary spacetime. In a stationary reference frame, there are two stationary observers, denoted by $G$ and $G'$, respectively. Let $\tilde{Z}^\mu$ represent the four-velocity of the observer and $\tilde{K}^\mu$ represent the four-wavevector of the photon, then, the energies of the photon relative to the stationary observer at points $p$ and $p'$ are $E|_p=-(\tilde{Z}^\mu\tilde{K}_\mu)|_p$ and $E|_{p'}=-(\tilde{Z}^\mu\tilde{K}_\mu)|_{p'}$, respectively.

Since the world line of the stationary observer coincides with the integral curve of the Killing field, $\tilde{\xi}^\mu=\zeta\tilde{Z}^\mu$, one can find $\zeta=(-\tilde{\xi}^\mu\tilde{\xi}_\mu)^{1/2}=(-g_{00})^{1/2}$, which leads to $E|_p=[-(\tilde{K}_\mu\tilde{\xi}^\mu)\zeta^{-1}]|_p$ and $E|_{p'}=[-(\tilde{K}_\mu\tilde{\xi}^\mu)\zeta^{-1}]|_{p'}$. Considering that $\tilde{K}_\mu\tilde{\xi}^\mu$ is a constant along geodesics, $(\tilde{K}_\mu\tilde{\xi}^\mu)|_p=(\tilde{K}_\mu\tilde{\xi}^\mu)|_{p'}$, therefore
\begin{align}
    1+z=\frac{\zeta|_{p'}}{\zeta|_p}=\frac{\sqrt{-\tilde g_{00}}|_{p'}}{\sqrt{-\tilde g_{00}}|_p}\,.
\end{align}
In the Newtonian limit, the redshift can be further expressed as
\begin{align}
    z&=\sqrt{\frac{1+2\tilde\Phi|_{p'}}{1+2\tilde\Phi|_{p}}}-1\approx[\Phi+\omega(\phi)]|_{p'}-[\Phi+\omega(\phi)]|_p\,.
\end{align}
Since the contribution of the Newtonian potential can be calculated exactly or offset by experimental design, we will focus only on the contribution of the scalar field
\begin{align}\label{eq:redshift}
    z_{\phi}=\omega(\phi)|_{p'}-\omega(\phi)|_{p}\,.
\end{align}

\section{Continuous modeling}

\subsection{Experimental design}

\begin{table*}[htbp!]
\caption{Parameters of the relevant materials. UHV stands for ultrahigh vacuum, XHV stands for extremely high vacuum, and IPM stands for interplanetary medium. The environmental size of osmium can be obtained in the mineral layer, the environmental size of water can be obtained in Lake Baikal, and the environmental size of XHV/IPM can be obtained in the universe.}
\label{tab:material}
\centering
\renewcommand{\arraystretch}{1.8}
{\scriptsize
\begin{tabular}{cccccccc}
\hline
\hline
Material & Osmium & Water & \red{Fiducial gas} & Dilute gas & UHV & XHV & IPM \\
\hline
Mean density $\rho_\mathrm{ave}$(kg/m$^3$) & $2.26\times10^4$ & $10^3$ & $4.2$ & $4.2\times10^{-3}$ & $4.2\times10^{-9}$ & $4.2\times10^{-15}$ & $4.2\times10^{-21}$ \\
Particle density $\rho_0$(kg/m$^3$)  & $-$ & $-$ & $10^3$ & $10^3$ & $10^3$ & $10^3$ & $10^3$\\
Particle size $R$(m) & $-$ & $-$ & $10^{-10}$ & $10^{-10}$ & $10^{-10}$ & $10^{-10}$ & $10^{-10}$ \\
Mean distance $L_0$(m) & $-$ & $-$ & $10^{-9}$ & $10^{-8}$ & $10^{-6}$ & $10^{-4}$ & $10^{-2}$ \\
Reduced size $\tilde{R}$ & $-$ & $-$ & $10^{-1}$ & $10^{-2}$ & $10^{-4}$ & $10^{-6}$ & $10^{-8}$ \\
Environmental size $R_\mathrm{eniv}$(m) & $10^{2}$ & $10^{3}$ & $\red{-}$ & $-$ & $-$ & $10^{7}$ & $10^{7}$ \\
\hline
\end{tabular}
}
\end{table*}

\begin{figure}[htbp]
    \centering
    \includegraphics[width=0.7\textwidth]{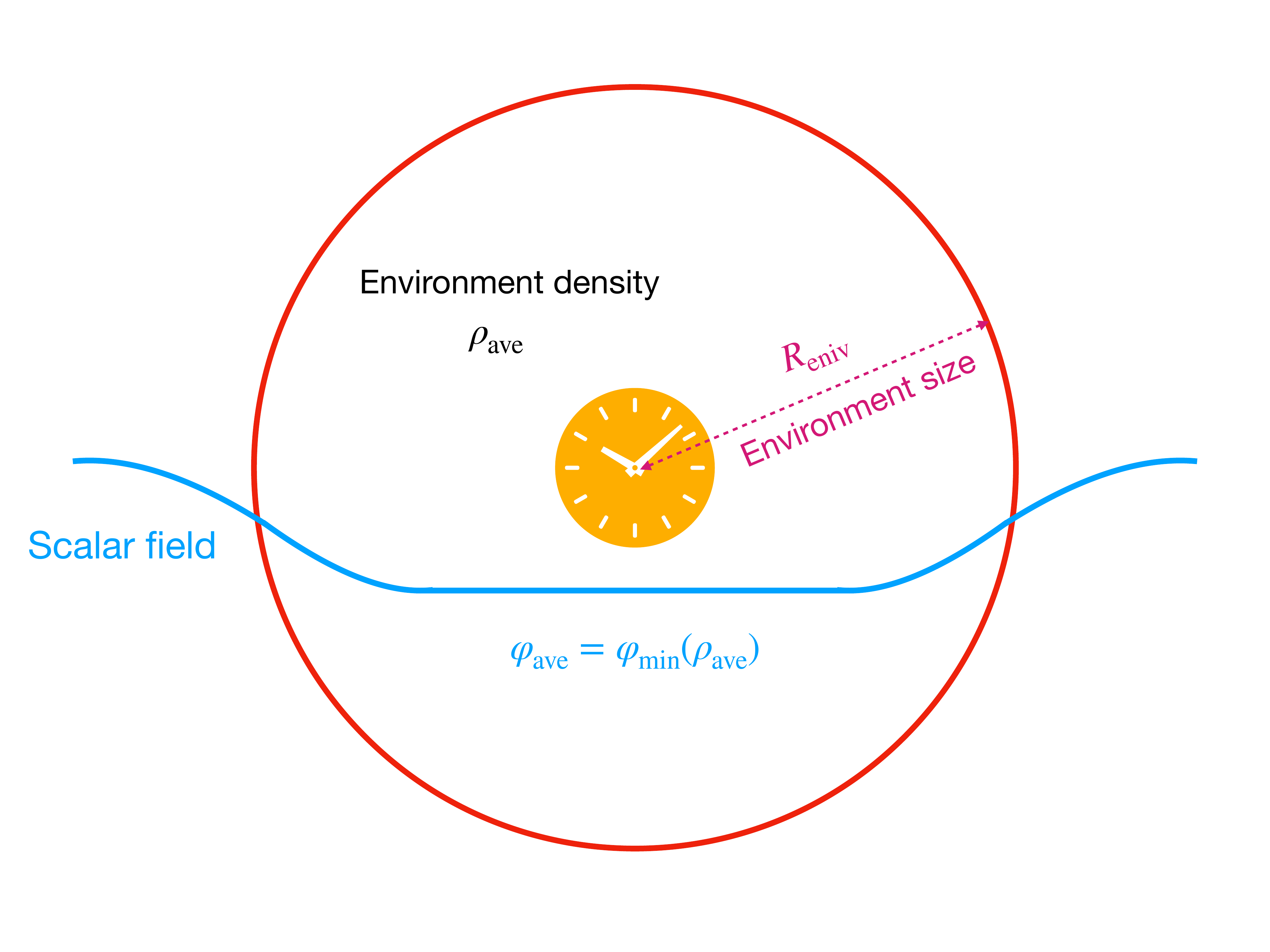}
    \caption{Gravitational redshift experimental scheme. The red circle represents the boundary of the vast environment. The blue curve shows the profile of the scalar field. }
    \label{fig:draw}
\end{figure}

We first study the case of a continuous mass distribution, i.e., a uniform environmental density $\rho_\mathrm{ave}$. For an idealistic experimental scheme, as shown in Fig.~\ref{fig:draw}. An atomic clock is placed in a uniform-density $\rho_\mathrm{ave}$ environment. This environment can be artificially constructed, such as by injecting substances of different densities into a vacuum chamber, or it can exist naturally, such as in a mine, lake, or outer space. For an idealistic experiment design, we first assume that the volume or characteristic size $R_\mathrm{eniv}$ of this environment is large enough so that the value of the scalar field $\varphi$ at the location of the atomic clock always corresponds to an extremum $\varphi_\mathrm{ave}$ of the effective potential. Then, we denote the frequency at which the atomic clock stably operates in a lower-density environment as $f$, and subsequently, the frequency at which the atomic clock stably operates in a higher-density environment as $f'$. At this point, we have completed the experimental measurements, and the redshift can be directly read from
\begin{align}
    z=f/f'-1\,.
\end{align}

The value of the scalar field is determined by the average density $\phi_\mathrm{ave}\equiv\phi_\mathrm{min}(\rho_\mathrm{ave})$, hence the corresponding redshift caused by the scalar field is obtained by combining Eq.~(\ref{eq:extremum}) and (\ref{eq:redshift}).
\begin{align}
    z_{\phi}&=\frac{A_2}{2M_\mathrm{Pl}^2}[(\phi_\mathrm{1,ave}-\phi_\ast)^2-(\phi_\mathrm{2,ave}-\phi_\ast)^2]\nonumber\\
    &=\frac{A_2}{2M_\mathrm{Pl}^2}[(\varphi_\mathrm{1,ave})^2-(\varphi_\mathrm{2,ave})^2]\nonumber\\
    &=\frac{V_0^2}{2A_2}\left[\left(\frac{1}{4V_0+\rho_\mathrm{1,ave}}\right)^2-\left(\frac{1}{4V_0+\rho_\mathrm{2,ave}}\right)^2\right]\,,
\end{align}
which is actually independent of $\phi_\ast$, and hence the only relevant free parameters are $A_2$ and $V_0$.

\begin{figure*}[htbp!]
    \centering
    \includegraphics[width=\textwidth]{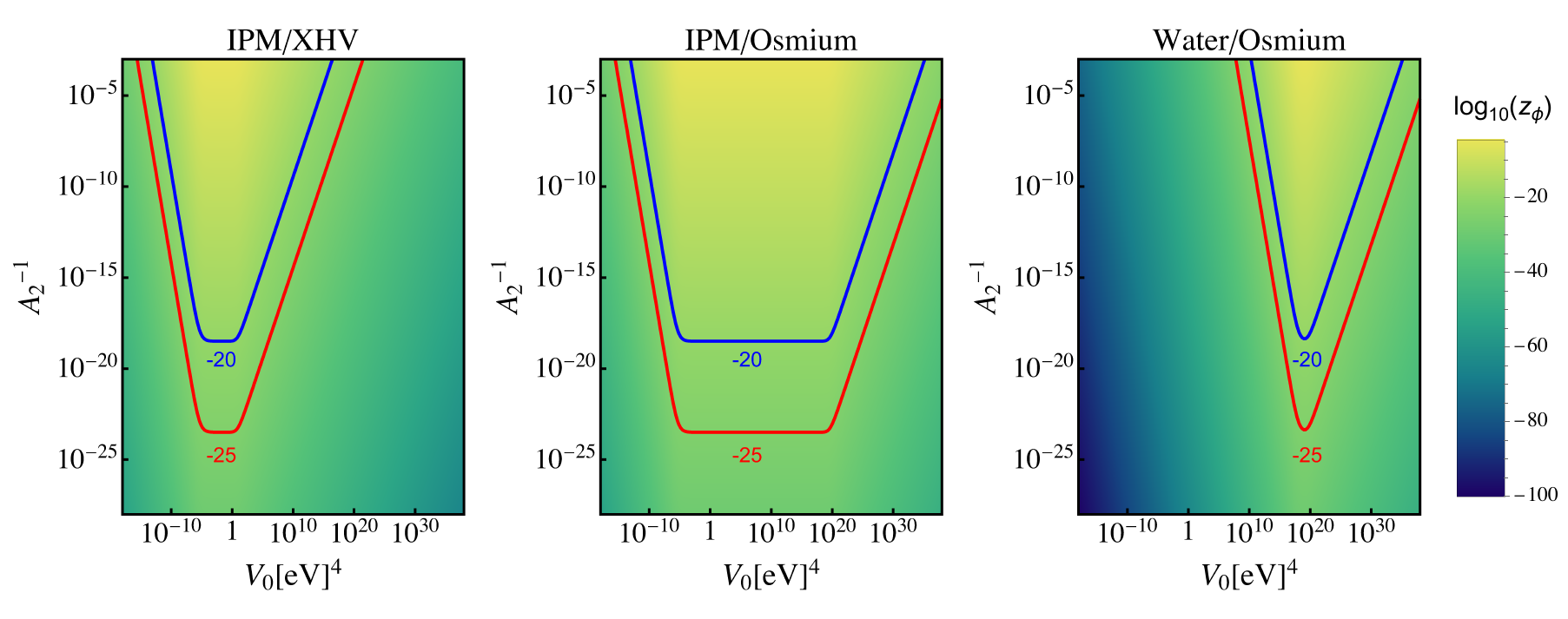}
    \caption{Expected values of gravitational redshift for different density pairs in a model with a continuous mass distribution. Note that redshift $z$ is independent of $\phi_{\ast}$ here. \red{The blue solid curve in the figure corresponds to a redshift value of $10^{-20}$, and the red solid curve corresponds to a redshift value of $10^{-25}$.} It can be found that the redshift decreases monotonically with the increase of $A_2$, but its relationship with $V_0$ is not monotonic.}
    \label{fig:ideal}
\end{figure*}

Here we list in Tab.~\ref{tab:material} the relevant parameters for different materials that are achievable with present-day technology, from the extremely low density of interplanetary medium (IPM) to the extremely high density of Osmium.

\subsection{Results and analysis}

\begin{figure*}[htbp]
    \centering
    \includegraphics[width=\textwidth]{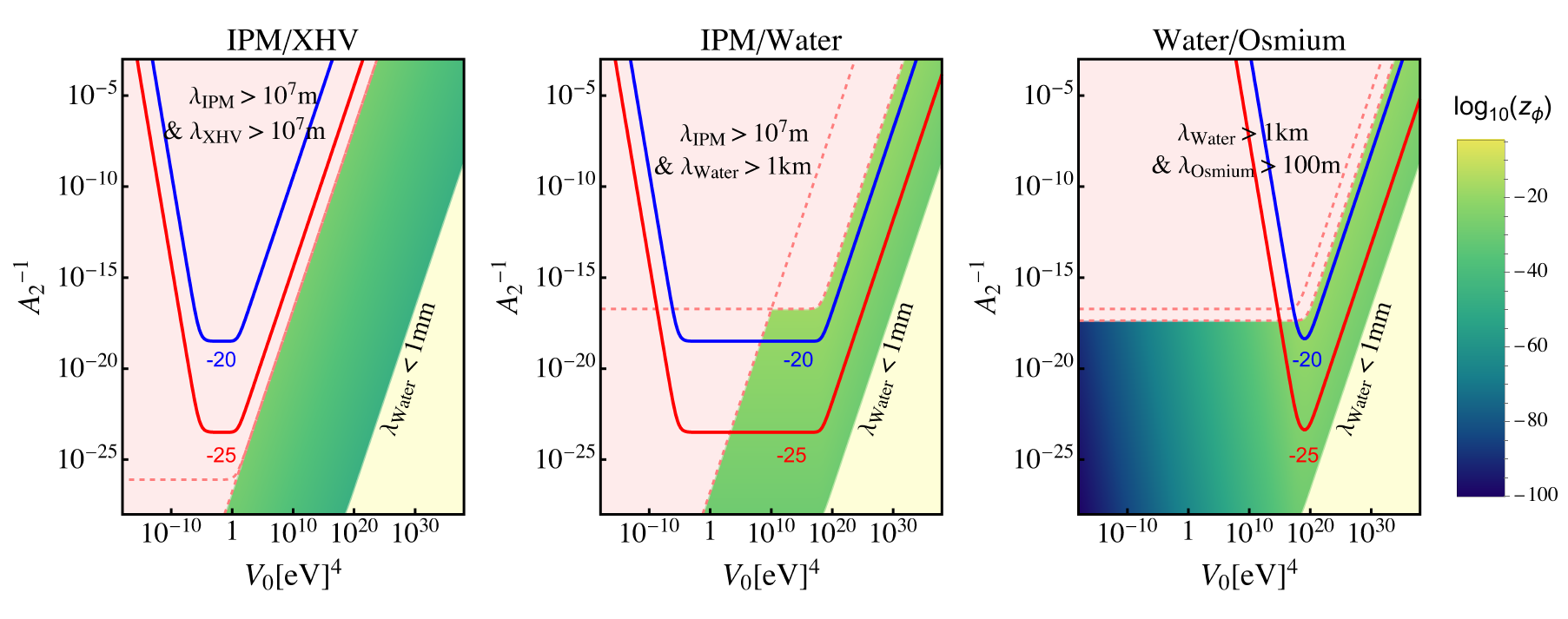}
    \caption{Expected values of gravitational redshift for different density pairs in the non-excluded region, \red{where $\lambda_\mathrm{material}\equiv\lambda_{\varphi}(\rho_\mathrm{material})$ represents the corresponding effective Compton wavelength.} The light yellow area represents the area that is excluded by the Compton wavelength of the \red{enclosure. Increasing the thickness of the protective enclosure would shift the light yellow area upward, thereby excluding a larger region.} The light red area represents the area that is excluded by the Compton wavelength of the finite-size environment. The light red \red{dashed} curves represent the Compton wavelength equal to the size of the environment. \red{The blue solid curve corresponds to a redshift of $10^{-20}$, and the red solid curve corresponds to a redshift of $10^{-25}$. Both of them lie partially or entirely within the excluded region.}}
    \label{fig:constraint}
\end{figure*}

In terms of theoretical predictions, the expected values of redshift contributed by the scalar field for three different density pair configurations are shown in Fig.~\ref{fig:ideal}. Here we compare three cases: (1) the two lowest density materials, (2) the lowest density material and the highest density material, (3) the two highest density materials. 

\red{Recent advances in timekeeping motivate our choice of reference redshift values. Arnold et al. have reported individual fractional systematic uncertainties near $1\times10^{-19}$~\cite{Arnold2026} for two $^{176}\mathrm{Lu}^{+}$ clocks. A specialized
optical lattice clock has achieved a fractional frequency uncertainty of $7.6\times10^{-21}$ \cite{Bothwell2022}. We therefore use $z=10^{-20}$ as an achievable benchmark, while the $10^{-25}$ contour is an ambitious benchmark, illustrating how a five-order improvement in the resolvable signal would change the parameter-space reach.}

It is straightforward to find that reducing the density of the lower one in the density pair can further push the constrained region to the left with lower energy scales of the dilaton potential; while increasing the density of the higher one in the density pair can further push the constrained region to the right with higher energy scales of the dilaton potential. In addition, comparing the lowest density to the highest density can obtain the maximally constrained region. Note also that, if we want to constrain the parameter $A_2$ to a larger range, the only way is to increase the precision of the atomic clock.

However, in the actual experimental design, we do not have an infinitely large environment, which means that the scalar field at the location of the atomic clock may not necessarily reach $\varphi_\mathrm{min}(\rho_\mathrm{ave})$. There is a simple criterion: if the Compton wavelength of the scalar field $\lambda_{\varphi}$ is less than the environmental size $R_\mathrm{eniv}$ of the corresponding substance, then the scalar field will be able to reach $\varphi_\mathrm{min}(\rho_\mathrm{ave})$ within this finite size.

Another challenge comes from the atomic clock itself. Typically, the atomic clock core is encased in a metal spherical \red{enclosure} to shield it from external environmental disturbances. However, for our gravitational redshift experiment, if the Compton wavelength reaches a value smaller than the thickness of the spherical \red{enclosure}, then no matter how the environment outside the \red{enclosure} is changed, the scalar field inside the spherical \red{enclosure} would not be affected, and this would make the experiment impossible to carry out!

In Fig.~\ref{fig:constraint}, we exclude these experimentally infeasible regions in the parameter space as indicated by light red and light yellow. For the characteristic size of the environment, we hope it can be as large as possible so that the detection region can be pushed more to the upper left. To be specific with the characteristic size, for IPM and XHV, the two density environments that are difficult to achieve on Earth, we can place the atomic clock along a distant retrograde orbit (DRO) via a rocket, with an orbital altitude of about 100,000 kilometers. Hence, it is pretty reasonable to estimate their characteristic size to be $10^7$ m. For water, the deepest lake on Earth, Lake Baikal, has a depth of up to 1 km. The average depths of the Pacific and Atlantic Oceans are both more than 3 km. Therefore, we set the characteristic size of water to 1 km. For osmium, we set the characteristic size of its mineral layer to 100 m. \red{For the metal spherical enclosure, we approximate its density by that of water; reducing the enclosure thickness shifts the infeasible region (light yellow) downward.}

Finally, it is easy to find that in the case of IPM/XHV, the detectable region has been completely excluded. In contrast, in the cases of IPM/Water and Water/Osmium, \red{there remain considerable regions that can be constrained using $z=10^{-20}$ and $z=10^{-25}$. The biggest problem is that it is difficult to impose any constraint on the large regions in the upper left corner of the parameter space. The reader may wonder, since the system is spherically symmetric, why not solve the EOM directly? The reason is that the Compton wavelength corresponding to the upper left region of the parameter space is too long for us to give an appropriate external boundary condition.} We emphasize that even so, the accessible region of gravitational redshift in the parameter space is still much higher than any of the regions excluded by existing tests, as the latter predominantly constrain the large-$A_2$ regime, while the former can probe smaller values of $A_2$.

\section{Discrete modeling}

The original process of obtaining a near-vacuum environment involves gradually extracting particles from the chamber. In this process, the volume of the chamber does not change, but the number of particles decreases, which is manifested as an increase in the average distance among particles. Then, there comes a natural question: when the environment density is low enough, does the assumption of continuous fluid still hold?

\subsection{Scalar field distribution}

\begin{figure}[htbp]
    \centering
    \includegraphics[width=0.7\textwidth]{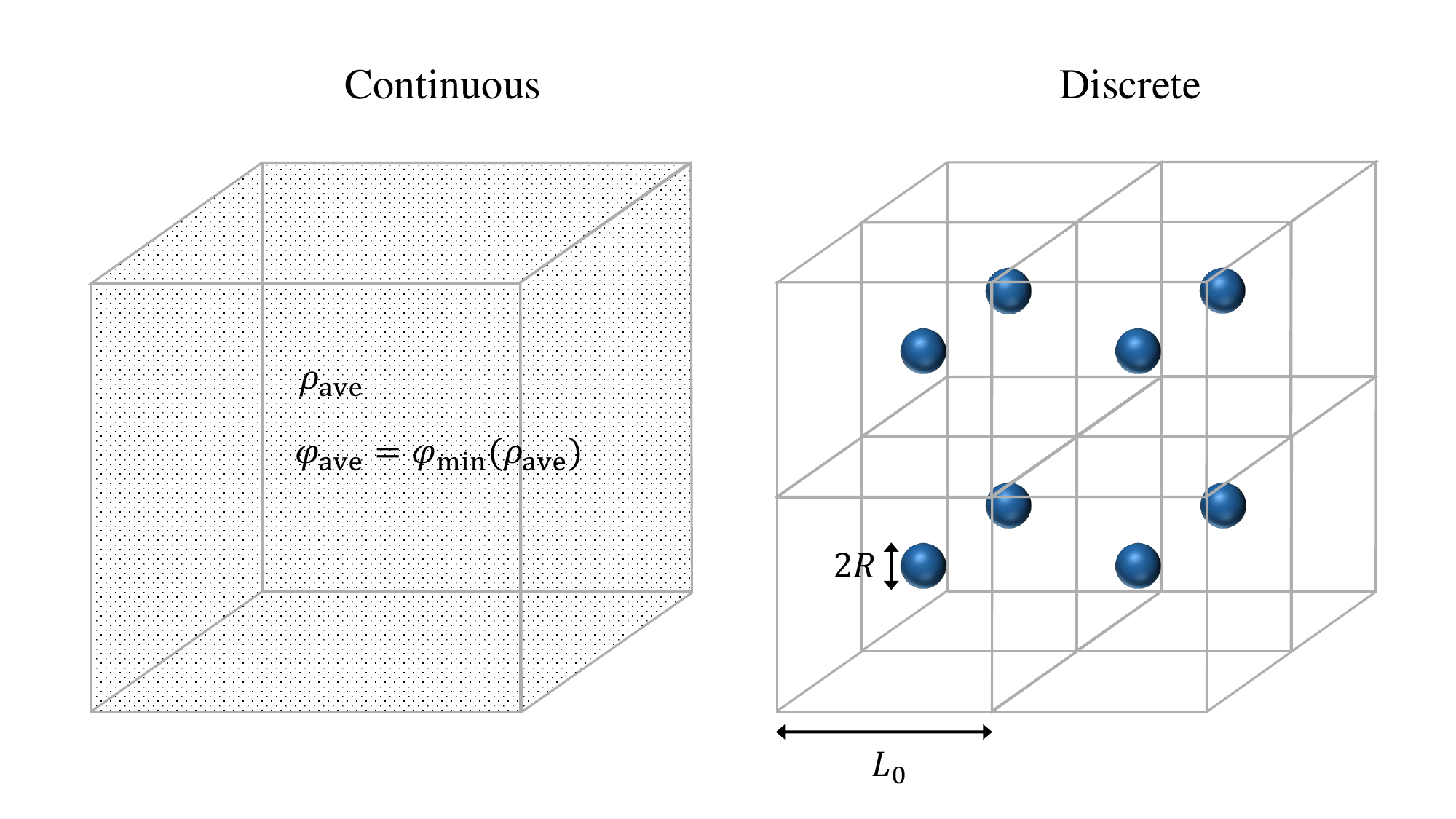}
    \caption{Continuous vs. discrete modelings. In the continuous model, the mass distribution of the matter field is uniform, so the density is constant, but in the discrete model, the mass within a cell is concentrated on a particle at the center, and the space between the particles is vacuum.}
    \label{fig:cubic}
\end{figure}

We study a model with a discrete mass distribution, whose mass is concentrated on spherical particles with a radius of $R$, and the space between particles is vacuum. This is inspired by the treatment of clumps of matter in cosmology~\cite{10.1111/j.1365-2966.2012.21750.x, Fleury:2016tsz, Fleury:2018cro}. We assume that the particles are periodically arranged as a lattice, as shown in Fig.~\ref{fig:cubic}. In this case, imagine that we use a cubic box with a side length $L_0$ to frame a particle and place it at the center of the box. The scalar field distribution inside the box can now be calculated directly since the boundary of the cube satisfies the periodic boundary conditions (symmetry ensures that the boundary conditions are also equivalent to the Neumann condition with no flux). 

Here we define a dimensionless spatial coordinate $\tilde{\mathbf{x}}=\mathbf{x}/L_0$, \red{the corresponding dimensionless Laplacian $(\tilde{\Delta}=L_0^2\Delta)$,} a dimensionless density $\tilde{\rho}=\rho/\rho_0$, and a dimensionless scalar field $\tilde{\varphi}=\varphi/\varphi_0$, where $L_0$ is the side length of the cubic box, $\rho_0$ is the density of the particle, and $\varphi_0\equiv\varphi_\mathrm{min}(\rho_0)$.

The specific values of different materials can be found in Tab.~\ref{tab:material}. It can be seen that we use a discrete model to treat the environment whose density does not exceed the density of air. In contrast, for the environment with density exceeding the air density, such as water or metal, the close packing between particles can be regarded as a continuous distribution of mass, so they are still treated with a continuous model. 

The dimensionless EOM can be expressed with dimensionless parameters $\eta=M_\mathrm{Pl}^2/(L_0^2A_2\rho_0)$ and $\xi=4V_0/\rho_0$,
\begin{align}
\eta\tilde{\Delta}\tilde{\varphi}=-(\xi+1)+\tilde{\varphi}(\xi+\tilde{\rho})\,.
\end{align}
Since the dimensionless density inside and outside the sphere is $1$ and $0$, respectively, the equations inside and outside the sphere can be expressed as
\begin{equation}\label{eq:neumann_eom}
\left\{
\begin{array}{cc}
     &\tilde{\Delta}\tilde{\varphi}-k_1^2\tilde{\varphi}=F\,,\quad\quad\red{0\le\tilde{r}\le\tilde{R}}  \\
     &\tilde{\Delta}\tilde{\varphi}-k_2^2\tilde{\varphi}=F\,,\quad\quad\red{\tilde{R}<\tilde{r}\le\tilde{a}}
\end{array}
\right.
\end{equation}
where $-F=k_1^2=(\xi+1)/\eta$, $k_2^2=\xi/\eta$\red{. $\tilde{R}$ is the reduced size of the particle, while $\tilde{a}=(\frac{3}{4\pi})^{1/3}$ is derived via an equal-volume approximation.} As detailed in~\ref{solve}, due to its approximate spherical symmetry, the EOM can be further simplified to the modified Bessel equation, whose solution is
\begin{equation}
\displaystyle
 \red{\tilde\varphi(\tilde r)=}
 \begin{cases}
 \red{C\,\dfrac{\sinh(k_1\tilde r)}{k_1\tilde r}
-\dfrac{F}{k_1^2}\,,}
&\red{0\leq\tilde r\leq\tilde R\,,}\\[6pt]
 \red{B\,\dfrac{\tilde a}{\tilde r}
 \left\{\cosh[k_2(\tilde a-\tilde r)]
 -\dfrac{\sinh[k_2(\tilde a-\tilde r)]}{k_2\tilde a}\right\}
-\dfrac{F}{k_2^2}\,,}
&\red{\tilde R<\tilde r\leq\tilde a\,.}
 \end{cases}
\end{equation}
\red{where the two coefficients $B$ and $C$ are defined in Eq.~\eqref{eq:neumann_coefficients}.}

With the profile of the scalar field, we can define the local average of the scalar field inside the box as
\begin{align}
\red{\left\langle\varphi(\mathbf{x})\right\rangle_{L_0}
\equiv
\frac{\int_\Omega\varphi(\mathbf{x})\mathrm{d}^3\mathbf{x}}{\int_\Omega\mathrm{d}^3\mathbf{x}}}=\varphi_0\left\langle\tilde{\varphi}(\tilde{\mathbf{x}})\right\rangle_1\,,
\end{align}
where $\Omega$ stands for the cubic box. 
\red{Then, using the equal-volume spherical-box approximation described in Appendix~\ref{solve}, the local average can be evaluated analytically,}
\begingroup
\color{black}
\begin{align}
    \varphi_0
    \left\langle\tilde{\varphi}(\tilde{\mathbf{x}})\right\rangle_1
    &\simeq
    \varphi_0
    \left\langle\tilde{\varphi}(\tilde r)\right\rangle_{\tilde a}
    =
    4\pi\varphi_0\int_0^{\tilde a}
    \tilde r^2\tilde{\varphi}(\tilde r)
    \,\mathrm{d}\tilde r
    \nonumber\\
    &=4\pi\varphi_0\Bigg\{
    C\left[
    \tilde R\cosh(k_1\tilde R)
    -\frac{\sinh(k_1\tilde R)}{k_1}
    \right]
    \left(\frac{1}{k_1^2}-\frac{1}{k_2^2}\right)
    \nonumber\\
    &\qquad
    -\frac{F}{3}
    \left[
    \frac{\tilde R^3}{k_1^2}
    +\frac{\tilde a^3-\tilde R^3}{k_2^2}
    \right]
    \Bigg\}
    \,.
\end{align}
\endgroup
Then, the redshift of the local average scalar field can be expressed as
\begin{align}
    z_{\phi}&=\frac{A_2}{2M_\mathrm{Pl}^2}[\left\langle\varphi_1(\mathbf{x})\right\rangle_{L_0}^2-\left\langle\varphi_2(\mathbf{x})\right\rangle_{L_0}^2]\,.
\end{align}

\subsection{Results and analysis}

\begin{figure*}[htbp]
    \centering
    \includegraphics[width=\textwidth]{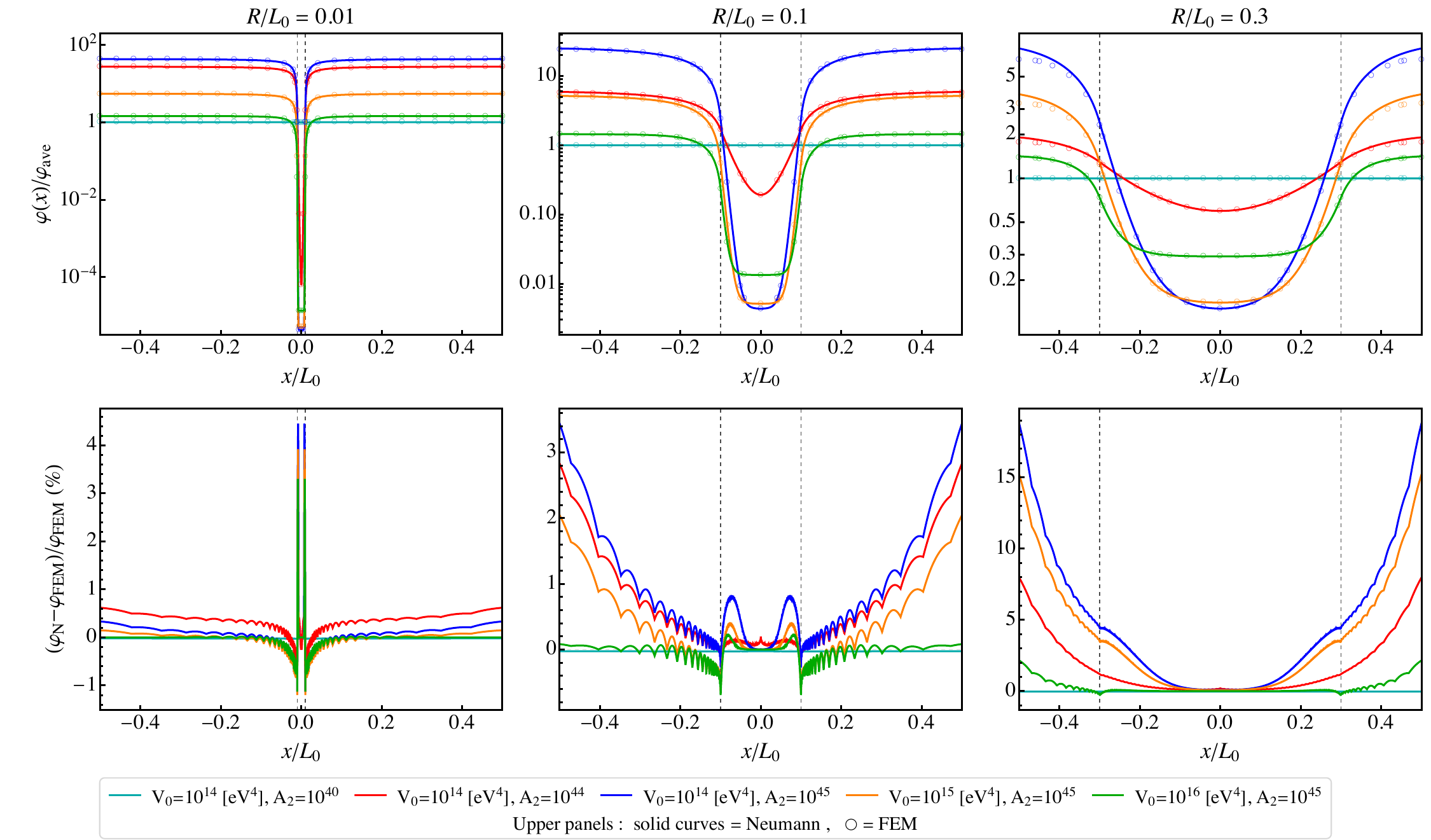}
    \caption{\red{The upper panels show the distribution of the scalar field $\varphi(\mathbf{x})$ on the x-axis in the cubic after normalization with $\varphi_{\mathrm{ave}}\equiv\varphi_{\mathrm{min}}(\rho_{\mathrm{ave}})=V_0M_{\mathrm{Pl}}/[A_2(4V_0+\rho_{\mathrm{ave}})]$. Here, we use the parameters of ``Fiducial gas'' given in Tab.~\ref{tab:material} as the benchmark parameters and keep the total mass of the central particle $\frac{4\pi}{3}\rho_0R^3$ fixed, change the particle radius $R$ and density $\rho_0$ at the same time. The reason is that the mass of microscopic particles is a directly measured quantity, but the density is an indirect quantity derived from mass and size, which depends on the accurate measurement of the particle size. Solid curves show the analytic Neumann solution, and open circles show samples from the three-dimensional finite-element solution on the original cubic box. The lower panels show the signed relative deviation $100[\varphi_{\mathrm{N}}(x)-\varphi_{\mathrm{FEM}}(x)]/ \varphi_{\mathrm{FEM}}(x)$ in percent. The two solutions agree well over most of the plotted domain. The profiles for $A_2=10^{40}$ are nearly uniform and close to unity, indicating that, in the cases shown, the scalar-field profiles depart appreciably from the continuous-model prediction only at stronger coupling (larger \(A_2\)).}}
    \label{fig:x_axis}
\end{figure*}

\red{In Fig.~\ref{fig:x_axis}, we provide the scalar field profiles on the $x$-axis of the cube under different parameter sets in the air density environment (middle). At the same time, we also studied the effect of increasing or decreasing the particle radius on the scalar field profile when the particle mass is fixed, because the mass of the particle is an accurate and directly measured quantity, while the radius and density are not. The gray dashed lines represent the boundary surface of the sphere. It is not hard to see that the scaling of both $V_0$ and $A_2$ will affect the scalar field distribution. And the ratio of the scalar field outside and inside the sphere decreases as the radius of the sphere increases (density decreases). Moreover, the most interesting finding is that when $A_2=10^{40}$, the scalar field distribution is uniform and indistinguishable from the values obtained in the continuous model, indicating that, in the cases presented, the scalar field distribution deviates significantly from the predictions of the continuous model only when the coupling is relatively stronger (i.e., when $A_2$ is relatively larger).}


\red{We also present finite-element results for comparison. In the upper panels, solid curves show the analytic Neumann solution, while open circles indicate the three-dimensional finite-element results obtained on the original cubic domain. The lower panels show the signed relative deviation, $100[\varphi_{\mathrm{N}}(x)-\varphi_{\mathrm{FEM}}(x)]/\varphi_{\mathrm{FEM}}(x)$, expressed as a percentage. The two methods are in good agreement.}

\begin{figure*}[htbp]
    \centering
    \includegraphics[width=\textwidth]{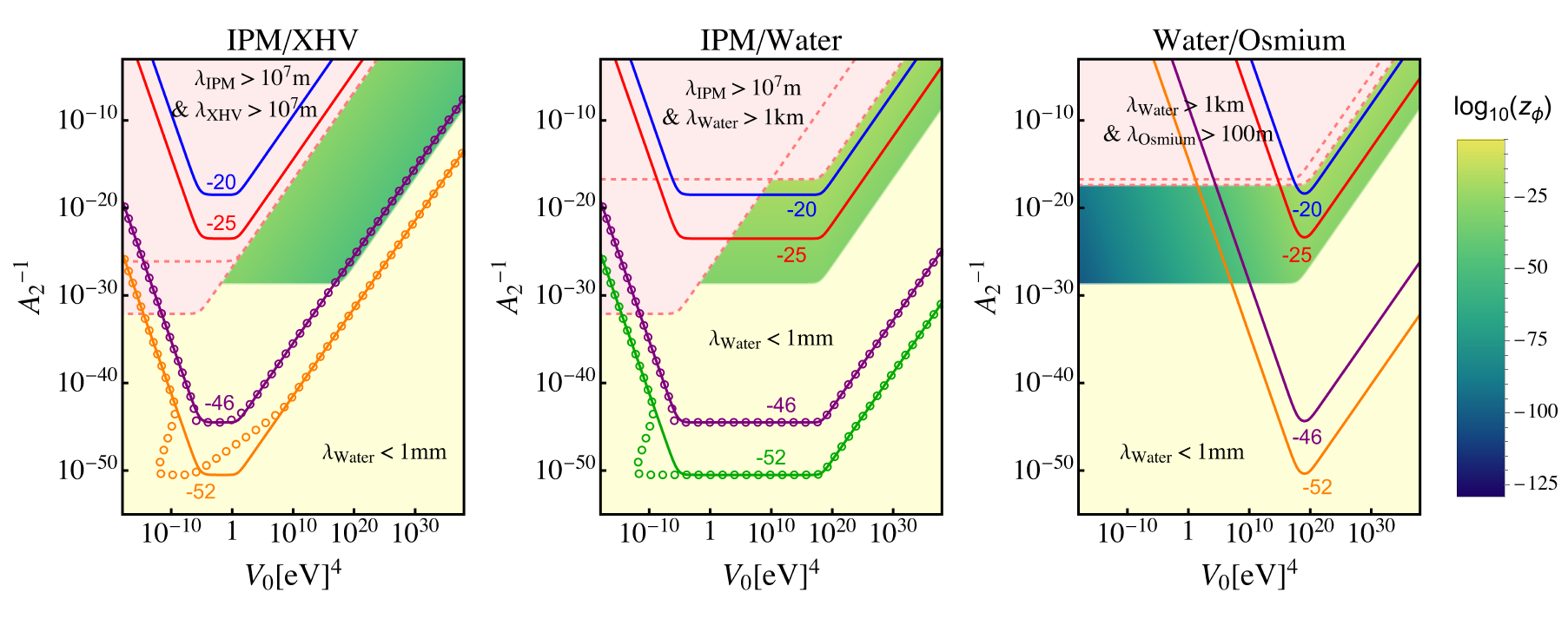}
    \caption{Expected values of gravitational redshift for different density pairs in a model with a discrete mass distribution, \red{where $\lambda_\mathrm{material}\equiv\lambda_{\varphi}(\rho_\mathrm{material})$ represents the corresponding effective Compton wavelength. The light yellow area represents the area that is excluded by the Compton wavelength of the enclosure. Increasing the thickness of the protective enclosure would shift the light yellow area upward, thereby excluding a larger region. The light red area represents the area that is excluded by the Compton wavelength of the finite-size environment. The light red dashed curves represent the Compton wavelength equal to the size of the environment. The solid curves represent the expected redshifts in the continuous model, while the curves with circular markers represent those in the discrete model. The two predictions become distinguishable only at a redshift measurement precision of $10^{-46}$ or better.}}
    \label{fig:average}
\end{figure*}

\red{We show the gravitational redshift for three different density pairs in Fig.~\ref{fig:average}. 
Here, the solid curves show the expected values of gravitational redshift in the continuous model, while the curves with circular markers
represent those in the discrete model. The deviations between the two predictions occur in the regime of very strong coupling, where the scalar field is already fully screened by the atomic clock's protective enclosure, rendering the field at the clock's core insensitive to changes in the external environment. Moreover, resolving these deviations in detection would require a redshift measurement precision of approximately $10^{-46}$ or better, well beyond what is expected to be achievable in the foreseeable future.}

\red{By comparing the discrete and continuous models, we find that they yield indistinguishable predictions within the parameter region accessible to current timekeeping technology, with their detectable regions coinciding exactly. Our calculations thus not only validate the continuous approximation in this regime but also clarify why it breaks down at relatively stronger coupling.}

\section{Conclusions and discussions}

This paper analyzes the possible constraints that the upcoming high-precision gravitational redshift experiment may impose on the parameter space of the dilaton.

We first consider all density environments modeled with a continuous mass distribution. In the continuous model, we study the expected value of redshift under ideal experimental conditions and find that increasing any one of them in the density pair can push the constraint region to the right side of the parameter space (larger $V_0$), and reducing any one of them in the density pair can push the constraint region to the left side of the parameter space (smaller $V_0$). Subsequently, we utilize the Compton wavelength to analyze perturbations in the outer protective spherical \red{enclosure} of the atomic clock and in the finite-size density environments. We find that it is unfeasible to detect some areas in the upper left and lower right corners of the parameter space under real experimental conditions. However, there are still quite large regions that high-precision gravitational redshift experiments in the foreseeable future can rule out.

\red{Our discrete calculation recovers the continuous-model prediction to the resolution relevant to the redshift values.
The small-scale departure from a homogeneous field appears at much stronger coupling in the cases studied, where the net scalar redshift is far below the timekeeping sensitivities, and the clock enclosure excludes this region. 
The agreement of the bulk predictions supports using a continuous-density model in the clock-relevant parameter region.}

\red{Under the stated timekeeping requirement, IPM/XHV has no bulk window at either benchmark, while comparisons involving water or osmium leave some sensitivity at smaller $A_2$.
Such clock comparisons could complement fifth-force searches, highlighting the unique role of gravitational redshift as a probe of weak-coupling regimes.}

\red{The redshift contours presented in our paper are based on somewhat idealized environments of representative sizes and a simplified model of the clock enclosure. Assessing the sensitivity of a specific experiment will require modeling its actual geometry and quantifying the uncertainty of the frequency comparison. In addition, the comparison with existing dilaton constraints is incomplete. As experimental data become available, bounds from clock comparisons could be plotted alongside existing limits in the same \((V_0,A_2)\) plane, making it possible to identify where redshift measurements offer complementary sensitivity.}

\begin{acknowledgments}
We thank Cheng-Gang Qin for helpful discussions.
This work is supported by the National Key Research and Development Program of China Grant No. 2021YFA0718304, No. 2021YFC2203004, and No. 2020YFC2201501, 
the National Natural Science Foundation of China Grants No. 12422502, No. 12105344, No. 12235019, No. 12447101, No. 12073088, No. 12575069, No. 11821505, No. 11991052, No.12475053, 12235016, and No. 11947302, 
the China Manned Space Program Grant No. CMS-CSST-2025-A01, 
the International Partnership Program of Chinese Academy of Sciences, No. 025GJHZ2023106GC.
the sub-project funding for ``Gravitational Redshift Measurement Scientific Experiment and Frontier Research in Gravitational Physics'' of the Chinese Academy of Sciences,
the Strategic Priority Research Program on Space Science, the Chinese Academy of Sciences (XDA30040000, XDA30030000)
\end{acknowledgments}

\appendix

\begingroup
\color{black}

\section{Derivation of scalar field profile}\label{solve}

We approximate this cubic box by a spherical box of equal volume,
\begin{align}\label{eq:neumann_cell_boundary}
 \frac{4\pi}{3}\tilde a^3=1\,,
 \qquad \tilde a=\left(\frac{3}{4\pi}\right)^{1/3}\,,
 \qquad \left.\frac{\mathrm d\tilde\varphi}
 {\mathrm d\tilde r}\right|_{\tilde r=\tilde a}=0\,.
\end{align}
The resulting radial solution is exact for this spherical box, while its use for the cubic lattice remains a geometric approximation. 

For a spherical profile, Eq.~\eqref{eq:neumann_eom} becomes
\begin{align} \label{eq:neumann_radial}
 \frac{\mathrm d^2\tilde\varphi}{\mathrm d\tilde r^2}
 +\frac{2}{\tilde r}
  \frac{\mathrm d\tilde\varphi}{\mathrm d\tilde r}
 -k_i^2\tilde\varphi=F\,,\qquad\qquad i=1,2
\end{align}
whose particular solution is $-F/k_i^2$. Subsequently, introducing $\tilde u=\tilde r(\tilde\varphi+F/k_i^2)$ for the homogeneous part gives $\tilde u''=k_i^2\tilde u$, where primes denote derivatives with respect to $\tilde r$.
Regularity at the origin therefore yields
\begin{align}\label{eq:neumann_inner}
 \tilde\varphi_{\mathrm{in}}(\tilde r)
 =C\,S(\tilde r)-\frac{F}{k_1^2}\,,
 \qquad S(\tilde r)=\frac{\sinh(k_1\tilde r)}{k_1\tilde r}\,,
 \qquad S(0)=1\,.
\end{align}

In the exterior region, both independent homogeneous solutions must be retained because the boundary is at finite $\tilde a$. Writing the homogeneous part as
\begin{align}
 \frac{\alpha\cosh[k_2(\tilde a-\tilde r)]
       +\beta\sinh[k_2(\tilde a-\tilde r)]}{\tilde r}\,,
\end{align}
the no-flux Neumann condition at $\tilde{r}=\tilde{a}$ gives
\begin{align}
\beta=-\frac{\alpha}{k_2\tilde a}\,.
\end{align}
Then, it is convenient to define
\begin{align}\label{eq:neumann_H}
 H(\tilde r)=\frac{\tilde a}{\tilde r}
 \left\{\cosh[k_2(\tilde a-\tilde r)]
 -\frac{\sinh[k_2(\tilde a-\tilde r)]}{k_2\tilde a}
 \right\}\,,
 \qquad H(\tilde a)=1\,,
 \qquad H'(\tilde a)=0\,.
\end{align}
The exterior solution is then
\begin{align}\label{eq:neumann_outer}
 \tilde\varphi_{\mathrm{out}}(\tilde r)
 =B\,H(\tilde r)-\frac{F}{k_2^2}\,.
\end{align}

The matching condition is that on the surface, $\tilde{r}=\tilde{R}$, the scalar field and its radial derivative must be continuous,
\begin{align}
 C S(\tilde R)-B H(\tilde R)
 &=F\left(\frac{1}{k_1^2}-\frac{1}{k_2^2}\right)\,,
 \label{eq:neumann_match_value}\\
 C S'(\tilde R)&=B H'(\tilde R)\,.
 \label{eq:neumann_match_derivative}
\end{align}
Introducing the notation $u=k_1\tilde R$, $v=k_2\tilde R$, $w=k_2\tilde a$, and define
\begin{align}
 S_R&\equiv S(\tilde R)=\frac{\sinh u}{u}\,,
 & H_R&\equiv H(\tilde R)
 =\frac{w}{v}\left[\cosh(w-v)-\frac{\sinh(w-v)}{w}\right]\,,
 \label{eq:neumann_surface_basis}\\
 d&\equiv\frac{S'(\tilde R)}{S_R}
 =\frac{u\coth u-1}{\tilde R}\,,
 & e&\equiv-\frac{H'(\tilde R)}{H_R}
 =\frac{1}{\tilde R}
 +k_2\frac{w\sinh(w-v)-\cosh(w-v)}
 {w\cosh(w-v)-\sinh(w-v)}\,.
 \label{eq:neumann_logderivatives}
\end{align}
Solving Eqs.~\eqref{eq:neumann_match_value} and
\eqref{eq:neumann_match_derivative} gives
\begin{align}\label{eq:neumann_coefficients}
 C=\frac{eF(1/k_1^2-1/k_2^2)}{(d+e)S_R}\,,
 \qquad
 B=-\frac{dF(1/k_1^2-1/k_2^2)}{(d+e)H_R}\,.
\end{align}

\endgroup

\bibliography{biblatex}

\end{document}